\documentclass[3p,hidelinks]{elsarticle}

\usepackage{hyperref}
\usepackage{amssymb, amsmath,color}
\usepackage{CJK}
\usepackage{makecell, threeparttable}
\usepackage{graphicx}
\usepackage[table]{xcolor}
\usepackage{arydshln}

\journal{ }

\makeatletter
\def\ps@pprintTitle{%
   \let\@oddhead\@empty
   \let\@evenhead\@empty
   \let\@oddfoot\@empty
   \let\@evenfoot\@oddfoot
}
\makeatother

\begin{document}

\begin{frontmatter}

\title{Return to basics: Clustering of scientific literature using structural information}

\author[a,b]{Jinhyuk Yun}
\author[b]{Sejung Ahn}
\author[b]{June Young Lee}\cortext[correspondingauthor]{Corresponding author}\ead{road2you@kisti.re.kr}
\address[a]{Department of Smart Systems Software, Soongsil University, Seoul 06978, Korea}
\address[b]{Future Technology Analysis Center, Korea Institute of Science and Technology Information \\ 66 Hoegiro, Dongdaemun-gu, Seoul, 02456, Korea}

\begin{abstract}
Scholars frequently employ relatedness measures to estimate the similarity between two different items (e.g., documents, authors, and institutes). Such relatedness measures are commonly based on overlapping references (\textit{i.e.}, bibliographic coupling) or citations (\textit{i.e.}, co-citation) and can then be used with cluster analysis to find boundaries between research fields. Unfortunately, calculating a relatedness measure is challenging, especially for a large number of items, because the computational complexity is greater than linear. We propose an alternative method for identifying the research front that uses direct citation inspired by relatedness measures. Our novel approach simply replicates a node into two distinct nodes: a citing node and cited node. We then apply typical clustering methods to the modified network. Clusters of citing nodes should emulate those from the bibliographic coupling relatedness network, while clusters of cited nodes should act like those from the co-citation relatedness network. In validation tests, our proposed method demonstrated high levels of similarity with conventional relatedness-based methods. We also found that the clustering results of proposed method outperformed those of conventional relatedness-based measures regarding similarity with natural language processing--based classification.
\end{abstract}

\begin{keyword}
Clustering \sep Mapping \sep Bibliographic coupling \sep Co-citation \sep Relatedness \sep Bipartite network 
\end{keyword}

\end{frontmatter}


\section{Introduction}\label{sec:introduction}
Although it is difficult to consider cutting-edge technology, the mapping of scientific publications is still a major sub-field in information science \cite{Small1997, Morris2008}. A typical approach is to use citation relations between scientific items (\textit{i.e.}, papers, patents, and authors). The citation network can be understood as a directed graph between articles, so direct citation (DC) is the most straightforward way of tracing the citation linkage from one article to another \cite{Small1999}. The major motivation of citations is to review and give credit to prior studies \cite{Case2000}. However, citations of relevant sources can be missing for various reasons, such as information overload, search failure, and non-use policy of journals \cite{Wilson1995}. Although up-to-date information technology enables more articles to be accessed online, this also steeply increases the amount of information. Still, there is always the possibility of missing the linkage between two scientific items.

\begin{figure*}[t]
\centering
\includegraphics[width=\textwidth]{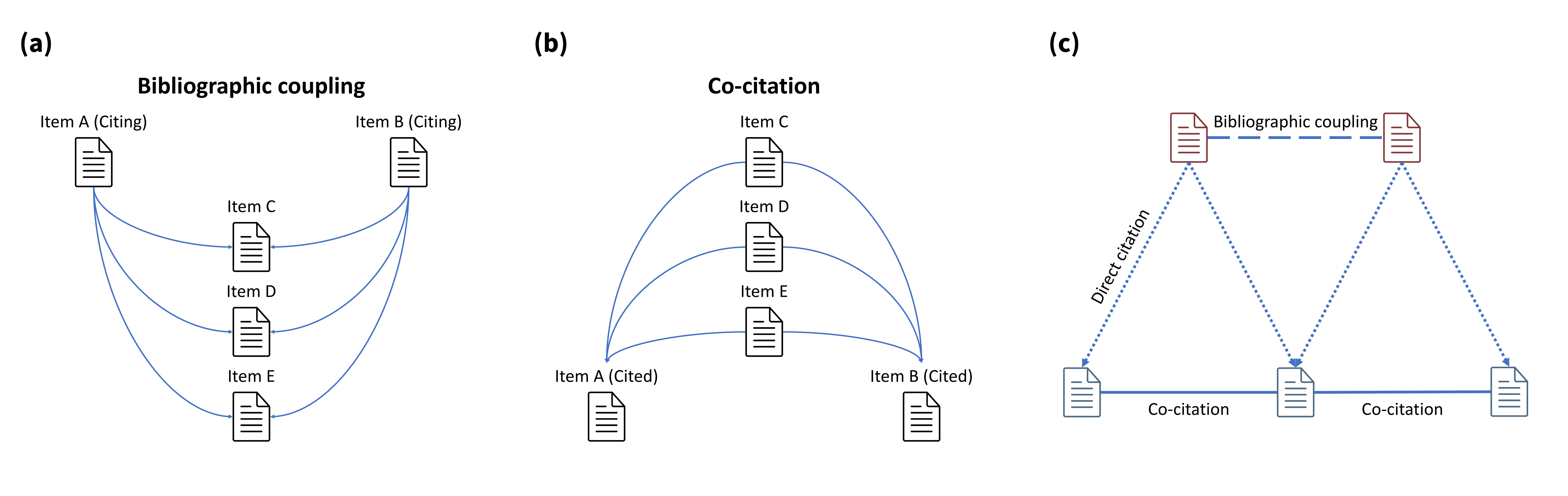}
\caption{Coupling methods: (\textbf{a}) BC links item A and B when they cite common items C, D, and E; (\textbf{b}) CC links items A and B when they are cited by common items C, D, and E. (\textbf{c}) The relationship between BC and CC can be interpreted as second-neighbor relations with inverse directions in a citation network.}
\label{fig01}
\end{figure*}

To estimate the missing links between two scientific items, scholars frequently employ relatedness measures. The two most common citation-based relatedness measures are coupling measures: bibliographic coupling (BC) and co-citation (CC). BC counts the relatedness between two scientific items when they have common third items in their reference lists~(Fig.~\ref{fig01}(a)) \cite{Kessler1963, Kessler1963a}. The strength of BC between two items is determined by the reference lists written by the original authors of the two items; thus, the linkage is based on the two authors' accumulated backgrounds when each article was published. Meanwhile, CC measures the frequency with which two items are cited together in a third item~(Fig.~\ref{fig01}(b)) \cite{Small1973}. CC conveys information on the cognitive similarity between two items from descendant authors because it depends on the reference lists of later publications. There are also many textual-based approaches introduced from the field of information retrieval such as BM25 and hybrid approaches that mix citation and textual similarity \cite{Liu2017, Waltman2020}.

Because of the numerous approaches in the field, the accuracy of each measure for mapping science and technology needs to be compared. Although no ground truth is available, some scholars have suggested determining the most accurate relatedness measure among the candidates \cite{Boyack2010, Klavans2017}. Other scholars have argued that the accuracy cannot be absolutely determined and that each relatedness measure is accurate from its own perspective, so finding one relatedness measure that is more precise than the others is unnecessary \cite{Glaser2017}. This camp's main claim is that different relatedness measures focus on different agents and viewpoints in science and technology. For example, the CC count varies over time, while the BC count does not change over time. We believe both views are correct in some sense, but the second approach suggests that each relatedness measure needs to be fitted to the purpose of the analysis. For a comprehensive understanding of science and technology, a specific relatedness measure is sometimes needed.

Modern advances in information technology enable a large amount of bibliographic metadata to be tracked. For example, Scopus contains more than 76 million records as of early 2020 and is rapidly increasing in size with time \cite{Bass2020}. Even though modern improvements in information technology have made available huge amounts of computational power, the number of records is also much greater than in the past. Calculating the relatedness of BC and CC is challenging because of the high computational complexity. The exact complexity depends on the implementation method but should be greater than linear. Thus, as new items are introduced into a database, more computational resources are needed to manage the extra data.

To overcome the above difficulties, we propose a new approach for identifying the research front that combines the principles of each relatedness measure, \textit{i.e.}, BC and CC. We applied a simple modification to the citation graph to emulate BC and CC without suffering from the massive computational complexity of relatedness measures. In particular, a simple node-splitting method for the DC graph makes it possible to get similar information as that of coupling relatedness measures. In this study, we mainly focused on the similarity between clusters from the original relatedness measures and those of our proposed method captured by an information entropy-based similarity measure. We also tested the performance when the citation linkages were normalized to find the optimal normalization for our proposed method. Our validation results showed that the clustering of the proposed method not only emulated the original relatedness-based clusters but also outperformed them in accuracy when compared to natural language processing-based article classification.

\begin{figure*}[h]
\centering
\includegraphics[width=\textwidth]{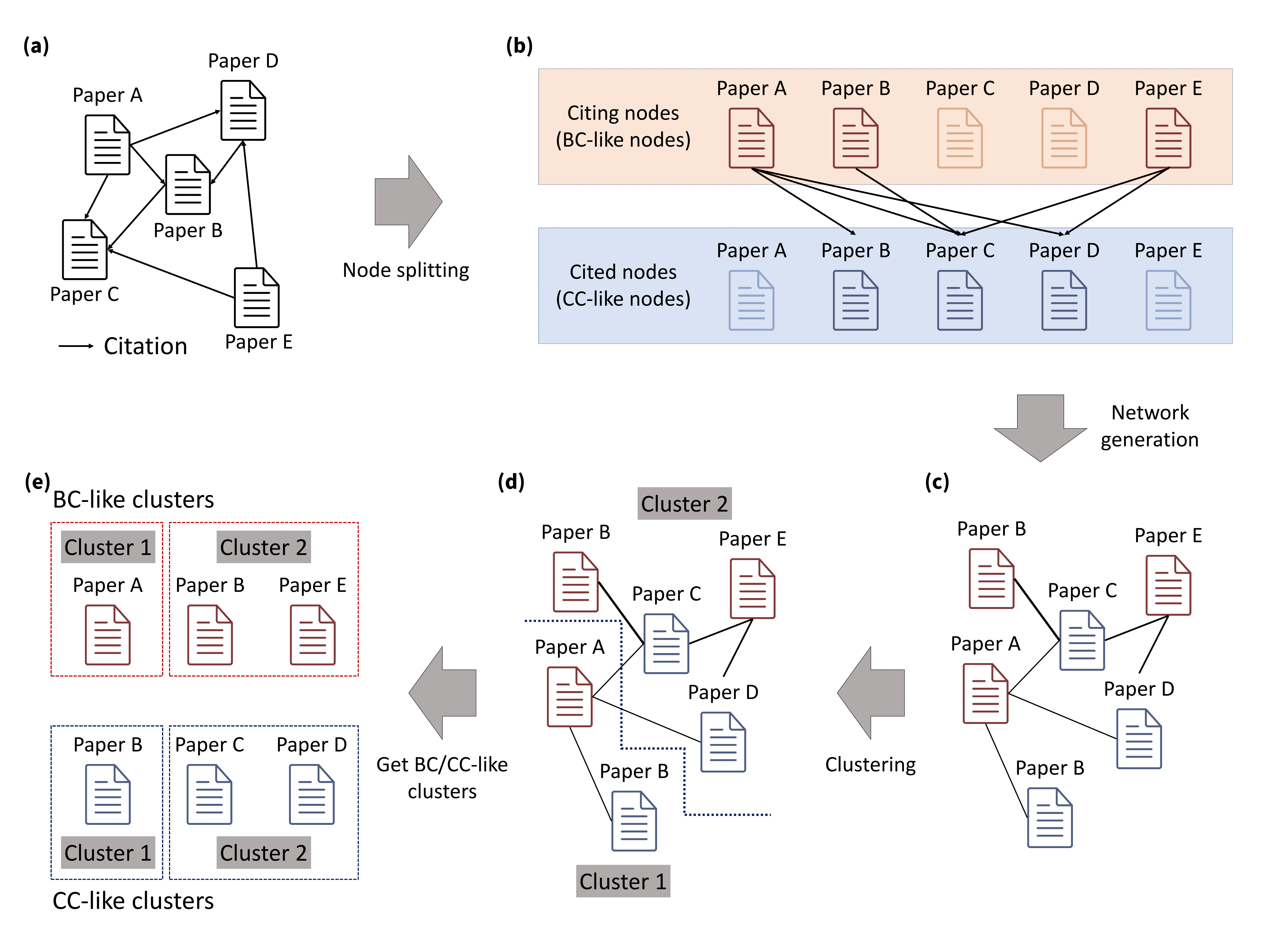}
\caption{Schematic diagram of the proposed clustering method. (\textbf{a}) A network comprising entire articles and citations is generated. (\textbf{b}) A node is split into two distinct nodes based on the citation direction. BC links items A and B when they cite common items C, D, and E, (\textbf{b}) whereas CC links items A and B when they are cited by common items C, D, and E. (\textbf{c}) The interrelation between BC and CC can be interpreted as second-neighbor relations in a citation network.}
\label{fig02}
\end{figure*}

\section{Methods}\label{sec:node_split}
\subsection{Node split method for graphing citations}
Citation relations can be understood with graph theory. Nodes represent scientific items (papers or patents), and the directed edges between two nodes represent citations. From this perspective, BC and CC can be interpreted as second-neighbor relations between two items~(Fig.~\ref{fig01}(c)). A bipartite network with two distinct types of nodes can be projected onto a single layer \cite{Zhou2007}. For instance, if cited items are nodes and citing items are virtual edges between them, this naturally yields information similar to CC by projection. This is also valid for BC in reverse. Thus, clustering in BC and CC networks is equivalent to clustering in the projected citation network. Going further, DC already has the BC and CC information as second-neighbor relations~(Fig.~\ref{fig01}(c)), so projection does not need to be employed. Instead, the clustering information of the layer simply needs to be taken separately, as long as the network is bipartite. However, the critical problem is that an empirical citation network is not bipartite. A single paper can cite another paper, while a third paper cites both of them; thus, a paper can act as both the cited node and citing node at the same time.

We propose the node split method, which allows citation networks to be considered as a bipartite network. For a citation network, the two main roles of a node are giving and receiving citations. Our idea is simply to split a single node into two distinct nodes based on its role. First, we begin with a raw directed network of papers and citations (Fig.~\ref{fig02}(a)). We then duplicate each paper $P(i)$ into two papers belonging to different layers: a citing node $P_o(i)$ and cited node $P_i(i)$ (Fig.~\ref{fig02}(b)). The two items in different layers are linked when a citation exists in the original network. As an illustrative example, if paper $P(A)$ cites another paper $P(B)$, there is an edge between nodes $P_o(A)$ and $P_i(B)$ in the node split network. Note that there is no edge between nodes in the same layer. We remove dangling nodes with no edge and convert a directed edge into an undirected edge to generate the final network (Fig.~\ref{fig02}(c)). Additionally, we may apply an edge weight normalization technique to enhance the clustering accuracy, which we discuss further in Section~\ref{sec:clustersim}. We then apply network clustering algorithms to the generated network. There is no restriction on algorithm selection, so any clustering algorithm can be applied \cite{Blondel2008, Traag2019, Rosvall2009}. The resulting clusters comprise both citing and cited nodes. We again separate the cited and citing nodes to yield two distinct clustering results. Because there is no edge between nodes in the same layer, the clusters in a single layer fully depend on the higher- and even-order neighbor relations. The second-neighbor relations should have the largest impact on the clustering. In other words, the resulting clusters of citing nodes $P_o(i)$ and cited nodes $P_i(i)$ emulate BC and CC clusters, respectively.

One of the most important advantages of the node split method is the computational cost compared with BC and CC. Although the computational complexity of clustering algorithms in time and space has not been well studied, the required resources essentially depend on the numbers of nodes and edges. Our proposed method does not change the number of edges and increases the number of nodes twofold at most compared to the original graph. In contrast, BC and CC drastically increase the number of edges. Although they do not change the number of nodes, the increased number of edges increases the spatial complexity. Thus, edge filtering techniques need to be applied to reduce the complexity of these coupling measures \cite{Waltman2020}, which the node split method does not need. Moreover, calculating the coupling relatedness is inherently costly. A typical approach is to multiply the adjacency matrix of the citation network. The CC adjacency matrix is given by $C=AA^T$, and the BC adjacency matrix is given by $B=A^TA$, where $A$ is an $m \times m$ DC adjacency matrix and $m$ is a number of nodes. The computational complexity of a general network is $O(m^3)$ at most for a dense matrix if the cost of the matrix transpose is neglected. Fortunately, the complexity can be up to $O(nnz(A) \times m)$ because the citation network is very sparse. However, this is still costly. Here, $nnz(A)$ is the number of nonzero elements (\textit{i.e.}, total number of citations of articles) for an adjacency matrix. In contrast, the node split method requires a single loop for the edges to re-index the citing and cited articles. The temporal complexity is therefore only $O(nnz(A))$. In summary, the node split method is more computationally efficient than coupling measures in terms of temporal and spatial complexities.

\begin{table*}[h]
\caption{Total number of nodes and edges for each benchmark set.}\label{table01}
\begin{center}
\begin{tabular}{lccc}
\hline
\thead{Source} & \thead{Nodes} & \thead{Nodes (without isolated)} & \thead{Edges} \\
\hline
Information Science & 13217 & 9945 & 63171 \\
Sociology & 1182674 & 352587 & 865051 \\
Materials Science & 1559740 & 862752 & 5457293 \\
Physics & 1739118 & 1182880 & 7806808 \\
\hline 
\end{tabular}
\end{center}
\end{table*}

\begin{table*}[h]
\caption{Total number of nodes in the sampled networks.}\label{table02}
\begin{center}
\begin{tabular}{lcccc}
\hline
\thead{Type of network} & \thead{Information Science} & \thead{Sociology} & \thead{Materials Sciences} & \thead{Physics}\\
\hline
DC & 9945 & 352587 & 862752 & 1182880 \\
DC (GCC) & 9779 & 318425 & 843300 & 1151450 \\
BC, Top 20 & 8132 & 204403 & 653917 & 820012 \\
BC, Top 20 (GCC) & 8060 & 195821 & 647521 & 807180 \\
CC, Top 20 & 7841 & 217343 & 646301 & 886303 \\
CC, Top 20 (GCC) & 7762 & 205913 & 639898 & 874018 \\
\hdashline
Node Split & 16344 & 460846 & 1332449 & 1763237 \\
Node Split (GCC) & 15903 & 404175 & 1295457 & 1699048 \\
Node Split, Citing & 8353 & 225449 & 673811 & 856792 \\
Node Split, Citing (GCC) & 8141 & 198262 & 655559 & 824999 \\
Node Split, Cited & 7991 & 235397 & 658638 & 906445 \\
Node Split, Cited (GCC) & 7762 & 205913 & 639898 & 874049 \\
\hline 
\end{tabular}
\end{center}
\end{table*}

\subsection{Constructing benchmark networks}\label{sec:benchmarknetwork}
Although the proposed method has the theoretical merit of computational efficiency, it needs to be empirically validated with real citation data. We constructed a benchmark for the proposed method from the October 2018 dump of Microsoft Academic Graph (MAG) \cite{Sinha2015, Wang2019}, which contains the complete entries of Microsoft Academic Service. This dataset includes metadata of items in \texttt{TSV} format; it includes articles and patents with the citation relations between them. The field of study (FOS) classified by MAG was assigned to all items. We first filtered papers published within 10 years from 2008 to 2017. We then collected papers in three different fields among 19 top-level FoSs for the benchmark: sociology, materials sciences, and physics. A set of articles published in three well-known information science journals between 1950 and 2017 was also selected: the \textit{Journal of Informetrics (JOI)}, \textit{Journal of the Association for Information Science and Technology (JASIST)}, and \textit{Scientometrics}. For all sets, we only used items for which \texttt{DocType} was assigned to \texttt{journal}.

We restricted our analysis on the citation relations so that both the citing and cited articles belonged to the target subset. Each set contained $9945$ nodes (information science) to $1182880$ nodes (physics without isolated nodes; see Table~\ref{table01}). We then constructed relatedness networks with the original DC, BC, and CC. Edges of coupling measures were filtered with the Top$M$ method at $M = 20$ based on coupling strength \cite{Waltman2020}. Here, the coupling strength was defined as the number of co-references (for BC) or co-citations (for CC) between two nodes. Only the giant connected component (GCC) of each network was considered, so nodes outside GCC were also filtered out (see the statistics in Table~\ref{table02}). Note that the number of nodes (articles) in the GCC was greater for the node split network than its counterpart (for BC, compare the nodes of BC and citing nodes of the node split GCC; for CC, compare the nodes of CC and cited nodes of the node split GCC). An important merit of the proposed method is that it does not remove edges according to their strength. Therefore, it can cluster more nodes than the coupling methods for analysis.

We also normalized relatedness measures by dividing the relatedness of a measure by the total sum of relatedness for a paper $P(i)$ \cite{Waltman2012}. Thus, the normalized edge weight $\hat{r}$ (\textit{i.e.}, DC, BC, or CC) of papers $P(i)$ to $P(j)$ is given by

\begin{equation}
\hat{r}_{ij} = \frac{r_{ij}}{\sum_k{r_{ik}}},
\end{equation}

\noindent where $r_{ij}$ is the raw relatedness strength between papers $P(i)$ and $P(j)$. This normalization rescales the edge strength for all papers to be at the same magnitude. Similarly, we tested four normalization methods for the node split network: the raw network (Raw), out-directional normalization (OutNorm), in-directional normalization (InNorm), and bidirectional normalization (BiNorm). All edges of the node split network must exist between a citing paper and cited paper, so the methods are distinguished by the basis of normalization. The out-directional normalization is performed on citing papers, so the out-directional normalized edge weight $\hat{w}^{o}_{ij}$ between a citing paper $P_o(i)$ and cited paper $P_i(j)$ is defined as

\begin{equation}
\hat{w}^{o}_{ij} = \frac{w_{ij}}{\sum_k{w_{ik}}},
\end{equation}

\noindent where $w_{ij}$ is the link strength between a citing paper $P_o(i)$ and cited paper $P_i(j)$. In-directional normalization is performed in the reverse direction of out-directional normalization on the cited papers, so the in-directional normalized edge weight $\hat{w}^{i}_{ij}$ between a citing paper $P_o(i)$ and cited paper $P_i(j)$ is defined as

\begin{equation}
\hat{w}^{o}_{ij} = \frac{w_{ij}}{\sum_k{w_{kj}}},
\end{equation}

\noindent Finally, bidirectional normalization considers both directions, so the normalized weight is defined as the geometric mean of the out- and in-directional normalized weights:

\begin{equation}
\hat{w}^{b}_{ij} = \sqrt{\hat{w}^{o}_{ij} \times \hat{w}^{o}_{ij}} = \frac{w_{ij}}{\sqrt{{\sum_a{w_{ia}}} \times \sum_b{w_{bj}}}}.
\end{equation}

\subsection{Clustering analysis of the benchmark networks}\label{sec:clustering}
To find the cluster for a given network, we applied the up-to-date Leiden algorithm \cite{Traag2019}, which is a refined version of the Louvain algorithm. The Leiden algorithm allows various modularity or quality functions to be used; we used the quality function of the Potts model with the configuration null model proposed by Reichardt and Bornholdt (RB model) \cite{Leicht2008}:

\begin{equation}
Q = \sum_{ij} \left(A_{ij} - \gamma \frac{k_i k_j}{2m} \right)\delta(\sigma_i, \sigma_j),
\end{equation}

\noindent where $A_{ij}$ is the edge weight between nodes $i$ and $j$, $k_i$ and $k_j$ are the node strengths of nodes $i$ and $j$, respectively, and $m$ is the total number of nodes. The resolution parameter $\gamma$ can be controlled to vary the number of clusters. Here, $\sigma_i$ denotes the assigned community of node $i$ where the delta function $\delta(\sigma_i, \sigma_j)$ is 1 if $\sigma_i$ = $\sigma_j$ and 0 for other cases.

\begin{figure*}[h!t]
\centering
\includegraphics[width=\textwidth]{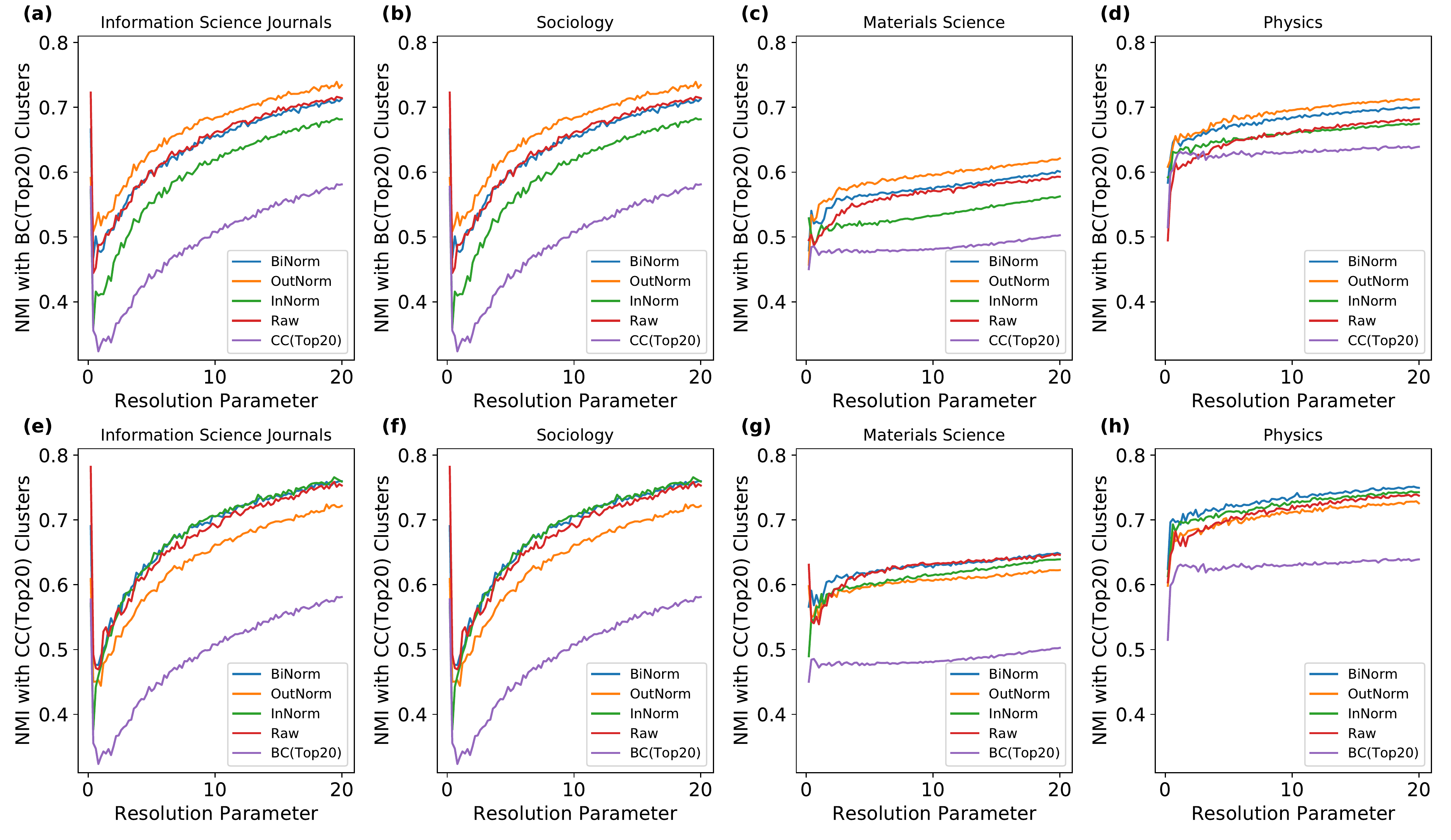}
\caption{Similarity between the clusters from the relatedness measures and node split network. (\textbf{a}--\textbf{d}) NMI similarity of clusters between BC(Top20) and the citing papers from the node split network with the four different normalization methods (Section~\ref{sec:clustering}). (\textbf{e}--\textbf{h}) NMI similarity of clusters between CC(Top20) and the cited papers from the node split network with the four different normalization methods (Section~\ref{sec:clustering}). For (\textbf{a}--\textbf{h}), the similarity between BC(Top20) and CC(Top20) is also presented to represent the baseline.}
\label{fig03}
\end{figure*}

We calculated the normalized mutual information (NMI) to estimate the similarity of the clustering results \cite{Danon2005, Gates2019}, which is based on information theory and is known for its robustness. Although NMI gives fair results for most cases, a bias has been reported for comparisons between different numbers of clusters \cite{Gates2019a, White1994}. Comparing clusters with similar granularity enhances the accuracy, so we considered the granularity of clusters for the results.

\section{Results}\label{sec:results}
\subsection{Similarity between clusters from the relatedness networks and node split network}\label{sec:clustersim}
Our main goal was to emulate the original relatedness clusters with a low computational cost. We used NMI to demonstrate the similarity between clusters from the relatedness networks and node split network. Our primary interest was to prove that our method provides results reasonably similar to the target clusters, not a perfectly match. The clusters from different methods should be similar to some degree because they reflect the landscape of science and technology. We compared our results with the coupling-based clusters (\textit{i.e.} CC and BC clusters) through NMI. Despite slight differences in disciplines, we found common high NMI scores for our proposed method (Fig.~\ref{fig03}). Specifically, NMI $>0.5$ for most of the range of the resolution parameter $\gamma$, which was much higher than the baseline similarity. We found that NMI was stable with varying $\gamma$ for large sets. NMI showed a large dip around $\gamma \sim 1.0$ for smaller subsets (Figs.~\ref{fig03}(a), (b), (e), (f)) but not for the large sets (Figs.~\ref{fig03}(c), (d), (g), (h)).

We continued our analysis by comparing the results with the normalization methods. Because the target relatedness network was normalized by the total sum of the relatedness of a given paper, normalization should increase the similarity. We indeed observed better similarity when the correct normalization was used. First, when BC was emulated for citing papers, out-directional normalization always outperformed the other three normalization methods (Figs.~\ref{fig03}(a)--(d)). However, when CC was emulated for cited papers, the results were rather unclear. Subsets of the information science and sociology journals gave better results with in-degree normalization (see Figs.~\ref{fig03}(e) and (f)). However, in-degree normalization was only the second-best method for physics (see Fig.~\ref{fig03}(h)) and third-best method for materials science (see Fig.~\ref{fig03}(g)). In both of the latter cases, bidirectional normalization outperformed the other normalization methods.

In summary, normalization increased the similarity between the original target clusters and node split clusters. Thus, for clusters of citing papers, we used out-directional normalization for the citing nodes and refer to this as bipartite backward citation coupling (BBCC). Despite the uncertainty in the results, we believe that in-directional normalization is the logical choice for clusters of cited papers if we consider the structure of the bipartite network in Fig.~\ref{fig02}. This is because we normalized the CC strength with the number of total co-citing articles, which is correlated with the number of in-directional links to cited nodes. Hence, we used in-directional normalization for the cited nodes, which we refer to as bipartite forward citation coupling (BFCC).

\begin{figure*}[h]
\centering
\includegraphics[width=\textwidth]{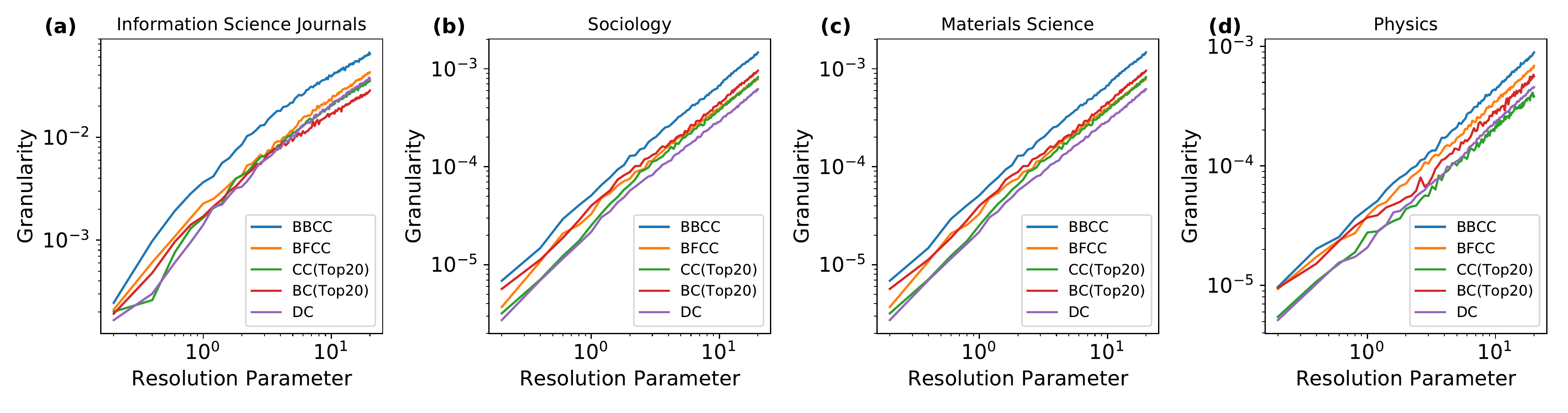}
\caption{Correlations between the granularity and resolution parameter for each network. Note that the granularity $G$ is an increasing function of the resolution parameter $\gamma$.}
\label{fig04}
\end{figure*}

\subsection{Clustering accuracy of the node split network compared to benchmark networks}\label{sec:benckmark}
Although it was not our primary interest, the general clustering similarity between the node split and coupling networks suggests an interesting question: Can the clustering results be improved with a measure other than relatedness? Obviously, there is no gold standard for perfect clustering. However, we believe textual-based similarity \cite{Waltman2020} is a good candidate. Thus, we again borrowed the FOS of MAG. This FOS encompasses natural language processing and citation-based similarity in a large-scale dataset that is organized hierarchically into six levels, where level 6 has the highest granularity. We used the level 2 FOS for validation data; this comprised $82,878$ different labels. Multiple FOSs were assigned to an article with a confidence score from 0 to 1. We did not allow the overlapping community structure at the clustering stage, so we only took the most confident label for each paper.

\begin{figure*}[h]
\centering
\includegraphics[width=\textwidth]{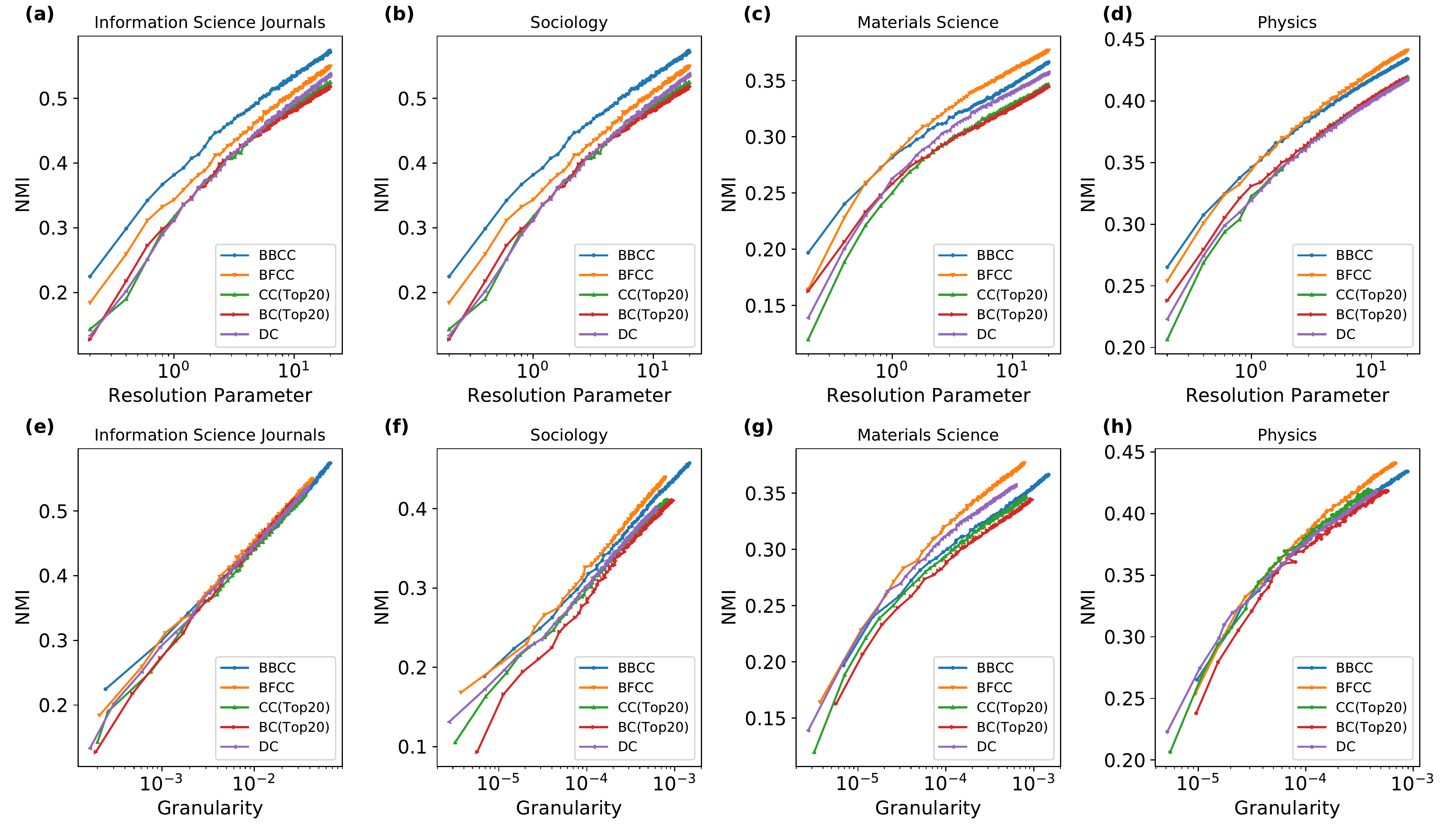}
\caption{Clustering accuracy for the $82,878$ clusters of the MAG Level 2 FOS according to NMI: (a--d) accuracy as a function of the resolution parameter $\gamma$ and (e--h) accuracy as a function of the granularity $G$.}
\label{fig05}
\end{figure*}

Before we proceed to the detailed analysis of the clustering accuracy, we stress that the granularity levels of each clustering method may differ, even for the same resolution parameter. The granularity $G^X$ for the clustering $X$ can be defined as

\begin{equation}
G^X = \frac{N}{\sum_{\alpha}{(S^X_{\alpha})^{2}}},
\end{equation}

\noindent where $N$ is the number of nodes in the set and $S_{\alpha}$ is the number of publications in the cluster $\alpha$ \cite{Waltman2020}. The granularity $G$ gradually increased as a function of $\gamma$, as we expected. However, the estimated granularity $G^X$ was quite different among the methods (see Fig.~\ref{fig04}). For the same $\gamma$, BBCC always had higher granularity than the other measures. To compensate for this effect, we evaluated the accuracy at each granularity level as well as resolution parameter value (Fig.~\ref{fig05}).

\begin{figure*}[p]
\centering
\includegraphics[width=0.8\textwidth]{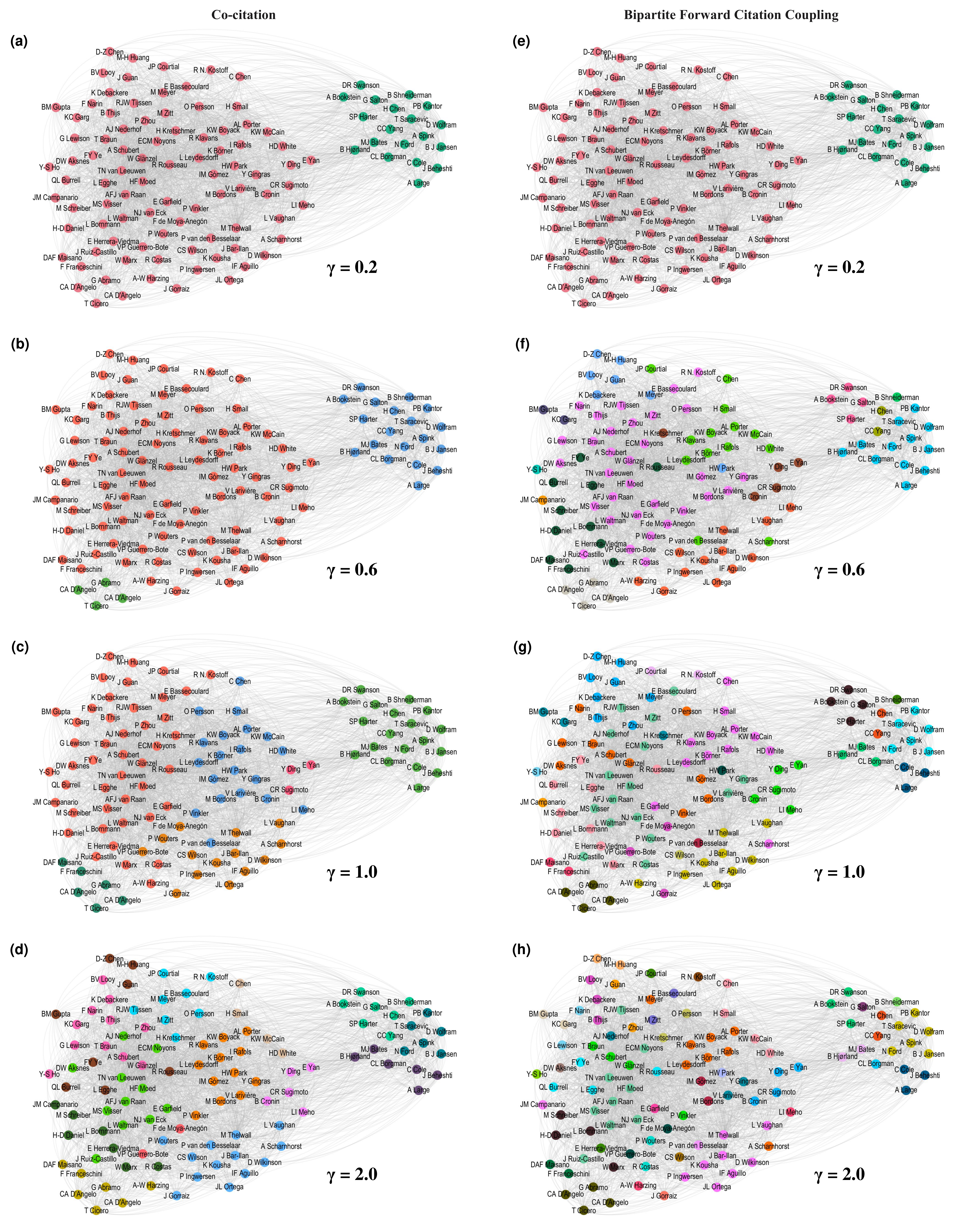}\label{fig06}
\caption{Application of the node split method to the author network. We selected the top 100 authors in the information science journals based on the $h$-index. Panels (\textbf{a}--\textbf{d}) show the clustering results with the CC network, and panels (\textbf{e}--\textbf{h}) show the results with the node split method. For (\textbf{a}--\textbf{h}), the edges represent CC relations, and the color corresponds to the clusters.}
\end{figure*}

In this analysis, we used NMI for the clustering results and MAG FOS to quantify the clustering performance of the different methods: a higher NMI indicated better clustering accuracy. First, the trend of NMI as an increasing function for both $\gamma$ and $G$ (see Fig.~\ref{fig05}) reflects its bias towards highly granular clusters \cite{Gates2019a, White1994}. Naturally, BFCC and BBCC outperformed the relatedness networks (BC, CC, and DC) for the same $\gamma$ level because they had higher granularity (see Fig.~\ref{fig04}). This trend was still valid when we accounted for the granularity. Both BBCC and BFCC had higher clustering accuracy than the corresponding BC and CC relatedness networks, respectively (see Fig.~\ref{fig05}(e--f)). Additionally, BFCC outperformed the other networks for most of the granularity range. Overall, the node split method not only reduces the computational cost but also may provide better clustering results.

\subsection{Application to coupling authors}\label{sec:application}
Although our clustering method is primarily intended for papers or patents, which cannot be expressed as a bipartite network, it can be applied to networks in a bipartite state. For instance, an author-paper citation network can be drawn as a bipartite network yet still be partitioned by our method. We demonstrate the results of the CC network and BFCC network for authors sampled from the information science field \cite{White1998}. We selected the top 100 authors in the set of information science journals presented in previous sections according to their $h$-index extracted from the citation in the set (see Fig.~\ref{fig06}. CC and BFCC gave similar results for low resolution parameter values ($\gamma \leq 0.2$), while BFCC gave more granular clusters for higher resolution parameter values ($\gamma \geq 0.6$). Besides the higher granularity, BFCC also yielded more singleton clusters than CC. These results demonstrate that the proposed method can be applied to any citation relations even if they are already in the bipartite state because it can yield similar clustering information as the corresponding relatedness network.

\section{Conclusion}\label{sec:conclusion}
In this study, we developed a novel approach towards mapping scientific literature that is inspired by the coupling relatedness measures of BC and CC. We applied our approach to an empirical dataset of papers in the fields of information science, sociology, materials science, and physics and demonstrated its merits. First, both layers of our method keep more nodes than its conventional coupling counterparts. According to NMI, our proposed method gives similar results without suffering from a heavy computational load. Level 2 FOS of MAG was used to evaluate the clustering accuracy, and the proposed method showed a higher clustering similarity than the coupling measures. Our validation was limited to certain fields of science within 10 years of publication, which certainly does not represent all fields of science and technology. The accuracy advantage may be smaller (or even negative) for other disciplines, but the proposed method still provides the advantage of a lower computational cost, which was our original objective. We also applied our method to clustering authors and compared the results with CC coupling to demonstrate its flexibility.

Although previous studies have examined relatedness measures in detail, they usually focused on calculating the relatedness directly, which requires a heavy computational load and high memory usage \cite{Small1997, Small1999, Kessler1963, Kessler1963a, Small1973, Boyack2010}. We suggest that an in-depth understanding of the citation structure may be needed to enhance the efficiency of the citation analysis. Our approach also can be understood as a refinement of investigating the extended direct citation \cite{Waltman2020}, which enhances the clustering accuracy with a minimum increment of the computational cost using structural information. Beyond the mapping harnessed in this study, many questions remain to be addressed. If data-driven analysis is based on a solid understanding of the mathematical formulation of the citation graph itself, the resulting synergy will help lead to an unbiased and quantitative understanding of the dynamics of science and technology.

\section*{Acknowledgments}
This work received institutional support from the Korea Institute of Science and Technology Information. This work was also supported by the National Research Foundation (NRF) of Korea funded by the Korean Government (Grant No. NRF-2017R1E1A1A03070975 (J.Y., S.A.)). The funders had no role in the study design, data collection and analysis, decision to publish, or preparation of the manuscript.

\bibliography{references}

\end{document}